\documentclass[fleqn,10pt]{article}

\usepackage{lipsum, ulem}
\usepackage{multicol}
\usepackage{placeins}
\usepackage{soul}
\usepackage{graphicx}

\begin{document}

\title{Composite material in the sea urchin \textit{Cidaris rugosa}: ordered and disordered micron-scale bicontinuous geometries\footnote{Manuscript accepted for publication in Journal of the Royal Society Interface, https://doi.org/10.1098/rsif.2023.0597}}

\author{
  Anna-Lee Jessop$^{1}$\footnote{Author for correspondence: Annie Jessop (Annie.Jessop@murdoch.edu.au)}, Allan J Millsteed$^{1}$, Jacob JK Kirkensgaard$^{3,4}$, Jeremy Shaw$^{2}$,\\ Peta L Clode$^{2,6}$, Gerd E Schr\"oder-Turk$^{1,5}$\footnote{$^{1}$School of Mathematics, Statistics, Chemistry and Physics, Murdoch University, Australia, $^{2}$Centre for Microscopy, Characterisation, and Analysis, University of Western Australia, Australia, $^{3}$Niels Bohr Institute, University of Copenhagen, Denmark, $^{4}$Department of Food Science, University of Copenhagen, Denmark, $^{5}$Research School of Physics, The Australian National University, Australia, $^{6}$ School of Biological Sciences, University of Western Australia, Australia}}


\maketitle


\section*{Abstract}
The sponge-like biomineralised calcite materials found in echinoderm skeletons are of interest in terms of both structure formation and biological function. Despite their crystalline atomic structure, they exhibit curved interfaces that have been related to known triply-periodic minimal surfaces. Here, we investigate the endoskeleton of the sea urchin \textit{Cidaris rugosa} that has long been known to form a microstructure related to the Primitive surface. Using X-ray tomography, we find that the endoskeleton is organised as a composite material consisting of domains of bicontinuous microstructures with different structural properties. We describe, for the first time, the co-occurrence of ordered Primitive and Diamond structures and of a disordered structure within a single skeletal plate. We show that these structures can be distinguished by structural properties including solid volume fraction, trabeculae width, and to a lesser extent, interface area and mean curvature. In doing so, we present a robust method that extracts interface areas and curvature integrals from voxelized datasets using the Steiner polynomial for parallel body volumes. We discuss these very large scale bicontinuous structures in the context of their function, formation, and evolution.

\section{Introduction}

Nature's ability to construct specialized composite materials through highly controlled biomineralization processes has inspired many decades of research (reviewed in \cite{weiner2008biomineralization, dove2018biomineralization}). Echinoderms, such as sea urchins and star fish, have been a focus of much of this research due to their ability to form skeletons with optimized physical properties \cite{yang2022damage, yang2022high, perricone2023echinoid} and the ease of imaging their optically clear embryos in real time (reviewed in \cite{wilt2007morphogenesis}). They form a hierarchical multi-element endoskeleton consisting of stroma and stereom: the stereom consists primarily of magnesium-calcite, small amounts of stable amorphous calcium carbonate, water, and intracrystalline organic molecules; the stroma consists of sclerocytes, collagen fibrils, and other extracellular matrix components \cite{gorzelak2021functional, smith1980stereom}. The stereom is formed through the biomineralization of calcium carbonate into magnesium-rich calcite that diffracts X-rays like a single crystal \cite{smith1990biomineralization, Nissen1969-ph, donnay1969x, yang2022damage}.


\begin{figure}[tbh]
    \centering
    \includegraphics[width=\textwidth]{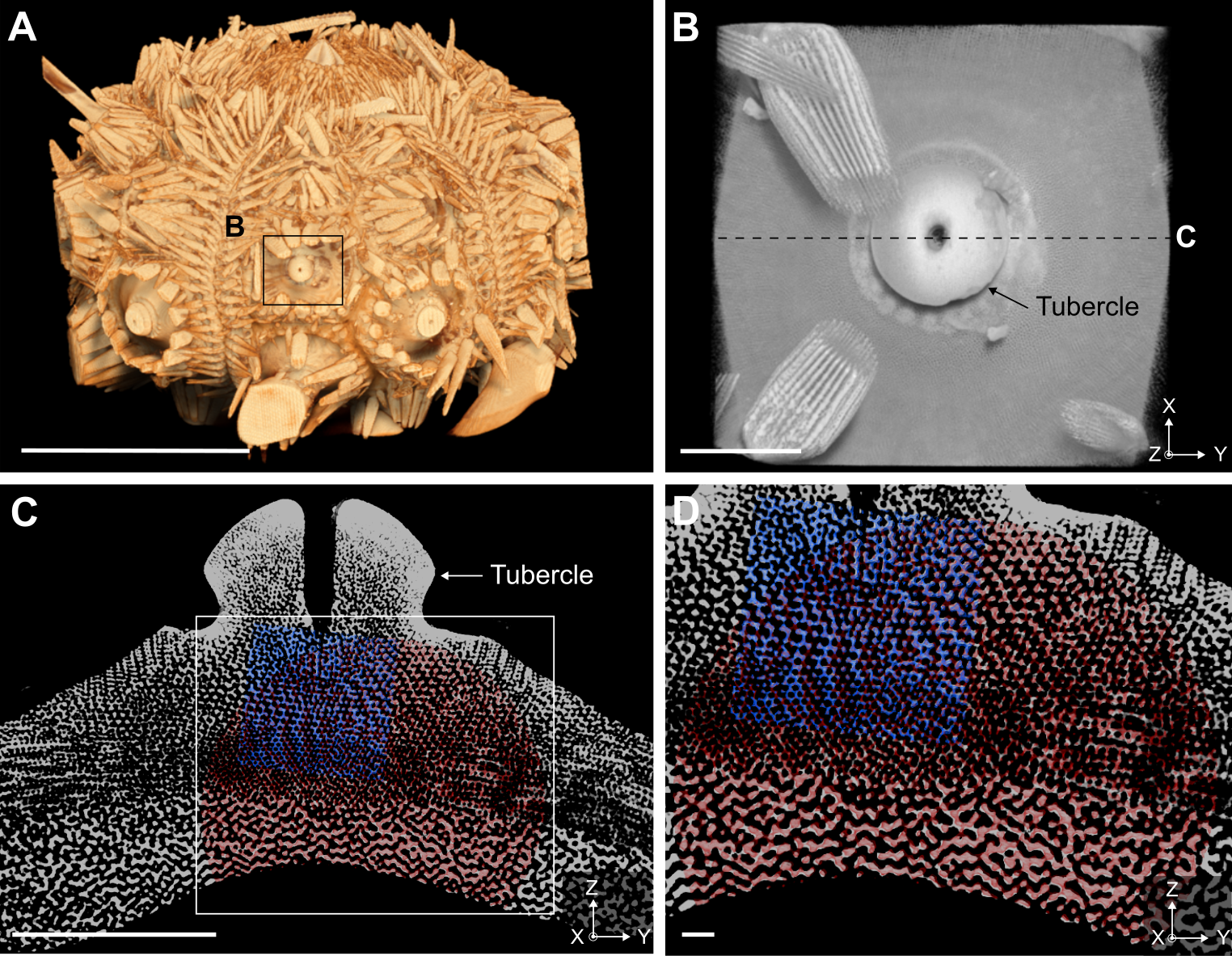}
    \caption{Reconstructed micro-CT data of the sea urchin \textit{C. rugosa}. (A) Three-dimensional visualization of the whole urchin skeleton depicting the approximate location of the sectioned interambulacral plate shown in B. The oral surface of the skeleton is pointing upwards and the aboral surface is pointing downwards. Scale bar = 1 cm. (B) Three-dimensional visualization of the sectioned interambulacral plate depicting the location of the slice shown in C. Scale bar = 1 mm. (C) A single slice through the interambulacral plate shown in C approximately oriented with the tubercle in the positive Z-direction. The outlined section depicts a close up of this slice shown in D. Scale bar = 1 mm. (D) Close up of outlined section in C, where the red highlight shows the region of the interambulacral plate that was scanned with a resolution of $1.69\,  \mu{m}$ and the blue highlight shows the region of the plate scanned with a resolution of 732 nm. Scale bar = $100\, \mu{m}$.}
    \label{fig:Figure1}
\end{figure}

Stereom types include dense solid layers of calcite and diverse three-dimensional mesh-works of trabeculae that exhibit complex sponge-like morphologies with curved surfaces \cite{smith1980stereom}. Many of these sponge-like stereom types can be thought of as bicontinuous geometries that partition space into two separate domains each of which forms a single connected network: one filled with stereom and the other with stroma. They can be disordered structures such as the 'laminar' or 'labyrinthic' stereom or they can exhibit a high degree of order such as in the 'rectilinear' stereom \cite{smith1980stereom}. In a few extraordinary cases, these ordered bicontinuous calcite structures have been shown to relate to known triply-periodic minimal surfaces (TPMS), albeit with solid volume fractions <50\% \cite{Nissen1969-ph, donnay1969x,yang2022damage, gorzelak2023devonian}. These structures exhibit periodicity in three spatial directions and partition space into two interconnected volumes, generating a porous light-weight structure with optimized properties such as high-energy absorption capabilities \cite{yang2022damage}. Remarkably, in all cases, the length scales (manifest in the lattice parameters) of these TPMS-related calcite structures are enormous (>$10 \mu{m}$; given they are thought to be templated by soft matter \cite{Hyde2022-xo}), being more than an order of magnitude larger than related structures found across most kingdoms of life (reviewed in \cite{han2018overview}). For example similar structures are found in insect cuticles and scales \cite{michielsen2008gyroid, saranathan2010structure, wilts2012brilliant,WiltsScienceAdv2017}, in intracellular membranes \cite{ALMSHERQI2009275,LUZZATI1997661, Kowalewska2019review}, and in the keratin nanostructures of bird feathers \cite{SaranathanBirdsPNAS}. Ordered bicontinuous structures related to TPMS geometries are hence common throughout the natural world, as they also are in physical chemistry (see reviews in \cite{LUZZATI200417,Hyde2009Elusive,MezzengaAdvMater2019}); however, the structures found in these echinoderms represent the only examples at such large length scales.

One of these examples was first described many decades ago in the sea urchin \textit{Cidaris rugosa} \cite{Nissen1969-ph}. Using scanning electron microscopy, Nissen (1969) described a stereom type within the interambulacral plates of \textit{C.\ rugosa} that exhibited a "regular network of trabeculae" and appeared to resemble one of the simplest TPMS, the so-called Primitive surface ('Plumber's nightmare'). Since then, authors have designated the 'rectilinear' stereom type, found in several species of Cidaroids \cite{smith1980stereom}, to the Primitive surface \cite{gorzelak2021functional, gorzelak2023devonian, ha2004three}, however a three-dimensional quantitative descriptive of this stereom type is lacking. Recently, a related ordered bicontinuous form, the so-called Diamond surface, was discovered in the ossicles of the starfish \textit{Protoreaster nodosus} \cite{yang2022damage} and in the fossilized skeleton of a 385 million year old crinoid \textit{Haplocrinites boitardi} \cite{gorzelak2023devonian}, sparking a renewed interest into the calcite microstructures of echinoderms including their potential in biomimetics \cite{perricone2023echinoid, perricone2022hexagonal, yang2022high, tian2023sea}. 

In the present study we re-examine the skeletal microstructures of \textit{C. rugosa} using X-ray tomography and X-ray scattering, providing a three-dimensional analysis of three of these microstructures and a preliminary analysis on their crystallinity. 
    
\section{Methods}
\subsection{Sample preparation}
A single specimen of \textit{C. rugosa} was donated by Prof. Charles Messing (Nova Southeastern University). This specimen had been collected during a manned submersible dive in the North Atlantic Ocean (26°07.1849’ N 79°50.0250’W) at a depth of 302 m in June 2006 and was identified by Dr David Pawson (Smithsonian Institute).

The spines of the whole urchin were removed and the whole urchin skeleton was glued to an aluminium mount and imaged. Subsequent to this, a single interambulacral plate was sectioned from the whole urchin using a 20 mm circular saw blade and glued to a pin for imaging.

\subsection{Image acquisition and processing}
Urchin samples were scanned with a Versa 520 XRM (Zeiss, Pleasanton, CA, USA) running Scout and Scan software (v15.0.17350.39816) at the Centre for Microscopy, Characterisation, and Analysis at the University of Western Australia. The whole urchin was scanned with a voltage of 65 kV at a magnification of 0.4X using the LE3 source filter delivering an isotropic voxel size of 22.8 $\mu$m. The sectioned interambulacral plate was scanned with a voltage of 80 kV at a magnification of 4X using the LE2 source filter delivering an isotropic voxel size of 3.72 $\mu$m. Two higher magnification scans were also conducted on an inner section of the interambulacral plate using a voltage of 80 kV, a magnfication of 4X, and the LE2 filter, delivering isotropic voxel sizes of 1.69 $\mu$m and 732 nm. Raw projection data were reconstructed using XRM Reconstructor software (v15.0.17350.39816) following a standard centre shift and beam hardening correction.

Three-dimensional micro-CT data were visualized and processed in Avizo 8.1 (Thermo Fisher Scientific, MA, USA), Dragonfly (v2022.1.0.1249, Object Research Systems, Montreal, Canada), and MATLAB 2021B (MathWorks®). The 16-bit grey scale intensity values of the inner interambulacral plate (isotropic voxel sizes of $1.69\, \mu{m}$ and 732 nm) had a clear bimodal distribution for values above zero. The grey scale data were converted to binary data by thresholding at the lowest frequency of the intensity distribution between the two modes. The 16-bit grey scale values of the whole interambulacral plate (isotropic voxel sizes of 3.72 $\mu$m) were first filtered using the three-dimensional 'Delineate' filter in Avizo 8.1 and then binarized using the same process described above. The binarized data set with isotropic voxel sizes of $1.69\, \mu{m}$ was used in all subsequent analyses.

\subsection{Structure identification}

Subvolumes of $94^3$ voxels corresponding to $(160\, \mu{m})^3$ resembling the $Fd\overline{3}m$ single Diamond structure and the $Pm\overline{3}m$ single Primitive structure were identified by visual inspection. To ensure the D-like and P-like stereom were in fact resemblances of these structures, nodal approximations with solid volume fractions and lattice parameters closely matching the structures were simulated and fit using manual registration in Dragonfly. 
Nodal approximations of the Primitive surface were modeled as level sets of the equation $\cos(2\pi x/a)+\cos(2\pi y/a) + \cos(2\pi z/a)=f$. Nodal approximations of the Diamond surface were modelled as level sets of $\phi_D(x,y,z)=f$ where
\begin{eqnarray}
\label{eq:diamond-nodal}
     \phi_D(x,y,z)&=&\sin\left(\frac{2\pi x}{a}\right)\sin\left(\frac{2\pi y}{a}\right)\sin\left(\frac{2\pi z}{a}\right)+\sin\left(\frac{2\pi x}{a}\right)\cos\left(\frac{2\pi y}{a}\right)\cos\left(\frac{2\pi z}{a}\right)\\&+&\cos\left(\frac{2\pi x}{a}\right)\sin\left(\frac{2\pi y}{a}\right)\cos\left(\frac{2\pi z}{a}\right)+\cos\left(\frac{2\pi x}{a}\right)\cos\left(\frac{2\pi y}{a}\right)\sin\left(\frac{2\pi z}{a}\right) \nonumber
\end{eqnarray}
    
The constant $f$ is adapted to give the desired volume fraction and $a$ is the lattice parameter. The common Miller index notation [hkl] is used to describe crystallographic directions; for example, the [100] direction are the coordinate axes (x,y,z) in eq. \( \ref{eq:diamond-nodal}\), and [111] the body diagonal (the notation is as in reference \cite{hain2023spire}).

The solid volume fraction was calculated for each subvolume as described below. The lattice parameter was approximated using the mean pore size (calculated as described below) of the subvolume relative to the mean pore size of a simulated nodal approximation with the same solid volume fraction and a lattice parameter equal to one.

In total, eight D-like stereom subvolumes were identified.
The $[111]$ direction of the Diamond structure (as shown in Figure \ref{fig:Figure3}B) was identified in each of subvolumes and the angle it forms with the spine direction (the positive Z-direction of our tomography data set) analysed.

We also analysed eight subvolumes with P-like stereom and eight subvolumes with disordered stereoms. For these, no orientation analysis was carried out.

\subsection{Distribution of solid volume fraction and pore sizes}

Our analyses are based on a segmentation of the tomography grey scale data into a binary data set consisting of voxels that have values of either one (part of the solid phase) or zero (part of the void phase). Mathematically, this is the discrete ('voxelised') analogue of the continuous indicator function $\chi(\mathbf{r})$ which equals one and zero for points $\mathbf{r}$ in the solid and void phase, respectively. We define the body representing the solid phase as $K=\{\mathbf{r}\,|\, \chi(\mathbf{r})=1\}$ and the void phase as $\overline{K}=\{\mathbf{r}\, | \, \chi(\mathbf{r})=0\}$

The solid volume fraction $\phi$ is the volume of the stereom (that is, the solid phase) divided by the overall sample size. A localised version of the solid volume fraction can be calculated for points across tomography data sets when dividing the sample into subvolumes and then calculating the same ratio for the binary data for each subvolume (for our main data set, we use subvolumes of $50^3$ voxels corresponding to $(84.5\, \mu{m})^3$). 

Pore size and trabeculae width distributions are defined based on the so-called 'covering radius transform' (CRT, \cite{Thovert,Mickel2008}) which in turn is derived from the Euclidean distance transform (EDM) which, for each voxel of a given phase, indicates the distance to the nearest voxel of the other phase. Mathematically, the Euclidean distance map is the function $D_{EDM}(\mathbf{r})$ defined for any point $\mathbf{r}\in \overline{K}$ in one phase as the distance to the nearest point of the other (which necessarily sits on the interface between $K$ and $\overline{K}$:
\begin{equation}
D_{EDM}(\mathbf{r})=\min_{(\mathbf{r'}\in K)}\left(|\mathbf{r}-\mathbf{r'}|\right).
\label{eq:edm-definition}
\end{equation}

The covering radius transform \cite{Thovert,Mickel2008} associates with each point $\mathbf{r}$ of the phase $K$, the value $D_{CRT}(\mathbf{r})$ of the largest EDM sphere that covers $\mathbf{r}$; an EDM sphere is a sphere centered at a point $\mathbf{r}'\in K$ with radius $D_{EDM}(\mathbf{r}')$. 

The distribution of the values of $D_{CRT}$ can be used as a meaningful pore size (providing the radii of the pores) or trabeculae width (providing the half-width of trabeculae) distributions of the body $K$.

\subsection{Surface area and mean curvature}

We have developed an algorithm, described in the Supplementary Materials, to estimate the surface area, the average mean curvature and the topological Euler characteristic. 
For a binary 3D data set, the algorithm computes the Euclidean distance function of a void phase. By counting the voxels with EDM $\le l$, it determines the volume $V_p(l)$ of the these sets for values of $l$ that are less than the so-called 'reach'. These sets are dilations (or parallel bodies) of the solid phase.
The so-called Steiner formula prescribes that $V_p(l)$ is a second-order polynomial in $l$  
\begin{equation}
\label{eq:steiner}
V_p(l)=(A l)\left(1+\langle H\rangle l+\frac{\langle G\rangle}{3} l^2\right)=A l +A \langle H\rangle l^2+\frac{2\pi \chi}{3} l^3
\end{equation}

where $A$ is the surface area of the solid-void interface $S$, $\langle H\rangle =\frac{1}{A}\int_S H(p)dA$ the average mean curvature of $S$ and $\langle G\rangle = \frac{1}{A}\int_S G(p)dA = \frac{2\pi \chi}{A}$ the average Gauss curvature of $S$ with the Euler characteristic $\chi$. 
The values of $A$, $\langle H\rangle$ and $\langle G\rangle$ can therefore be obtained by a fit to $V_p(l)$. For minimal surface model structures, we find that the fit produces accurate results for relatively low voxel discretisations; accuracy is further increased by combining data for $V_p(l)$ both the solid and the void phase. The algorithm works well for homogeneous structures where $l$ is relatively large. 

\subsection{Small- and Wide-angle X-ray Scattering (SAXS/WAXS)}

SAXS/WAXS measurements were performed using a Nano-inXider instrument from Xenocs SAS (Grenoble, France). The instrument is equipped with a Rigaku (Rigaku-Denki, Co., Tokyo, Japan) 40 W micro-focused Cu source producing X-rays with a wavelength of $\lambda$ = 1.54 Å. Scattered radiation is detected by two Pilatus detectors from Dectris (Baden, Switzerland) covering both SAXS and WAXS simultaneously. The two-dimensional scattering data were radially and azimuthally averaged using standard reduction software (XSACT, Xenocs). The scattering patterns for the radially averaged intensity were recorded as a function of the scattering vector q = 4$\pi$*sin$\theta$/$\lambda$ where 2$\theta$ is the scattering angle. Samples were mounted and measured in vacuum. Measuring times were 900 s. Samples were prepared by gentle drilling either from the external/internal side of the plate to create a bevelled region where, by small lateral adjustments of the X-ray beam, the sample thickness could be varied; measurements were ultimately conducted on samples approximately 0.5 mm thick which enabled appropriate flux.

\section{Results}

The endoskeleton of the sea urchin \textit{C. rugosa} has an oblate spheroid shape (major axis diameter = 22.2 mm) with pentaradial symmetry, dividing the skeleton into five interambulacral zones (comprising the interambulacral plates) and five ambulacral zones (comprising the ambulacral plates). Each interambulacral zone consists of between 10 and 11 interambulacral plates of varying sizes with the smallest occurring near the oral and aboral surfaces and the largest occurring through the centre of the skeleton. All interambulacral plates are covered in protrusions called tubercles (which house the spines of the sea urchin via a ball-and-socket joint \cite{takahashi1967ball}). The interambulacral plate examined here is located centrally on the skeleton and is approximately 3.7 mm in width (Figure\ \ref{fig:Figure1}A, B). It comprises diverse types of microstructures that are easily visualized at different resolutions both in cross-section and volume rendering ($3.72\, \mu{m}$, $1.69\, \mu{m}$, 732 nm; Figure\ \ref{fig:Figure1}C, D). An imperforate stereom is found around the external surface of the tubercle and the remaining microstructures are porous. The porous microstructures sampled here can be grouped into either ordered or disordered bicontinuous structures depending on the regularity of their trabeculae networks.

\subsection{Ordered stereom types resemble Diamond and Primitive bicontinuous geometries}

\begin{figure}[t]
    \centering
    \includegraphics[width=\textwidth]{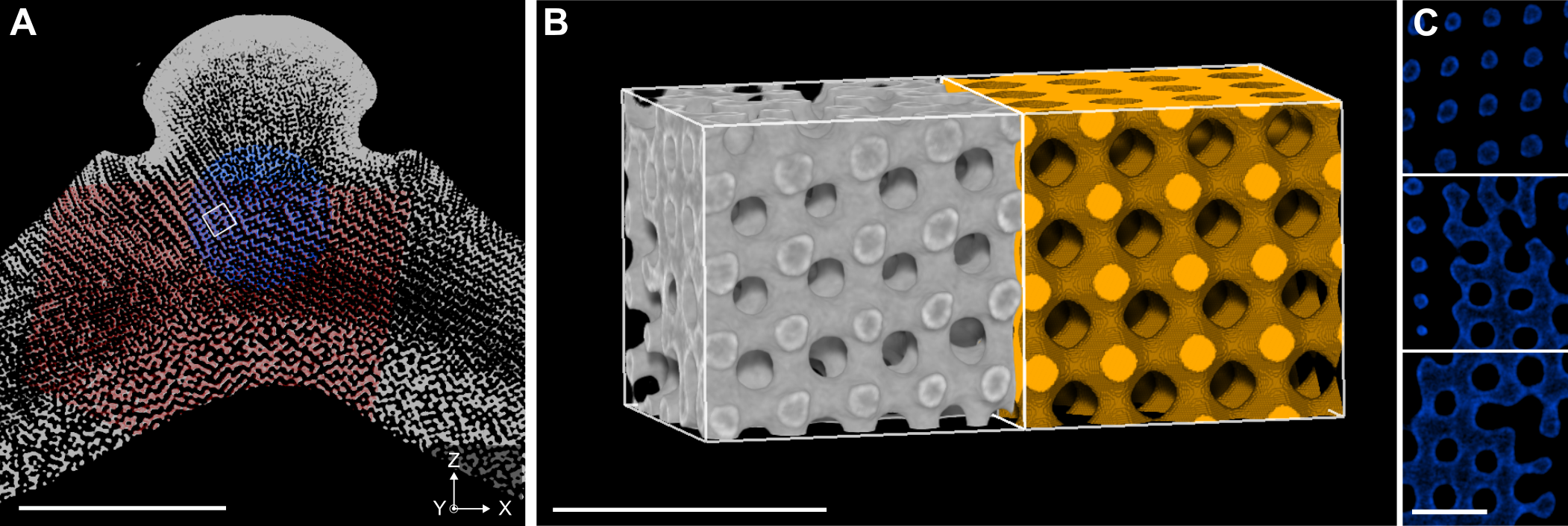}
    \caption{A representative subvolume of the stereom that closely resembles a single Primitive surface. (A) Cross-section through the interambulacral plate showing the location of the representative subvolume. Scale bar = 1 mm. (B) The subvolume of the sea urchin stereom (in grey; isotropic voxel size of 732 nm) and a simulated nodal approximation of the single Primitive surface with a solid volume fraction $\phi=0.38$ (in yellow) and lattice parameter $a=30$ microns. Scale bar = $100\, \mu{m}$.(C) Perpendicular slices through the sea urchin subvolume depicted in (B). Scale bar = $50\, \mu{m}$} 
    \label{fig:Figure2}
\end{figure}

We identified two ordered stereom types that both have a regular network of trabeculae but can be distinguished by their topological and geometric differences. These are based on companion surfaces of Schwarz Diamond minimal surface and Schwarz Primitive minimal surface. In line with Nissen's finding\cite{Nissen1969-ph}, we identified a stereom type that closely resembles the Primitive minimal surface geometry and is located in a layer of stereom that lies between the tubercle and approximately 1 mm below the tubercle (Figure\ \ref{fig:Figure2}). This P-like stereom is a slightly distorted (likely monoclinic or triclinic) version of the Primitive geometry with cubic $Pm\overline{3}m$ symmetry. Three translational repeat vectors can be identified in each sample with the $[100]$, $[010]$, and $[001]$ directions of the bicontinuous Primitive geometry. The average angle between these directions is 77\textdegree$\pm 4$\textdegree. Figure\ \ref{fig:Figure2}B displays a sample of the P-like stereom (in grey) such that the three orthogonal directions of the cube surrounding it roughly align with the $[100]$, $[010]$, and $[001]$ directions of the Primitive geometry. Adjacent to the P-like stereom is a nodal approximation of the Primitive geometry (in yellow) which, in the example in Figure\ \ref{fig:Figure2}B, has a lattice parameter equal to 30 microns and a solid volume fraction of 0.38. The lattice parameters of the P-like stereom samples, as estimated from the $\langle D_{CRT}\rangle$ value (Table\ \ref{tab:exp-PDG-area-curvature-results}), varied between 23 $\mu{m}$ and 25 $\mu{m}$. These estimates agree, up to approximately $\pm 1\, \mu{m}$, with estimates for the lattice parameter by manual registration of the nodal surface model.

\begin{figure}[tbh]
    \centering
    \includegraphics[width=\textwidth]{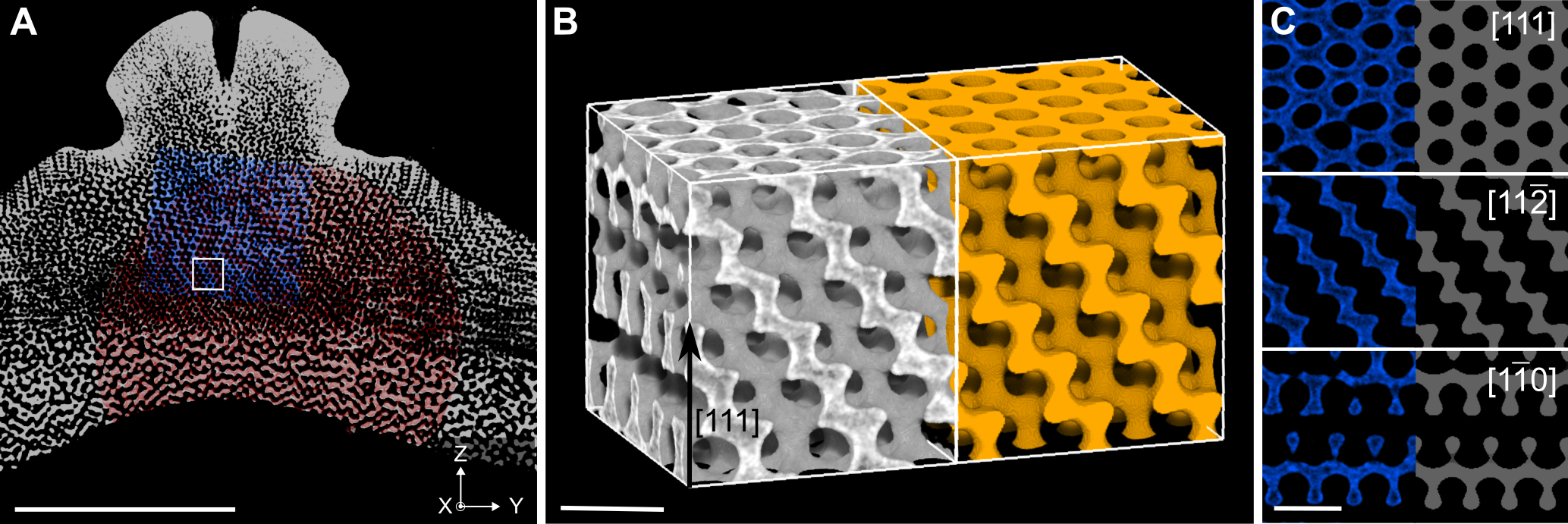}
    \caption{A representative subvolume of the stereom that closely resembles a single Diamond surface. (A) Cross-section through the interambulacral plate showing the location of the representative subvolume. Scale bar = 1 mm. (B) The subvolume of the sea urchin stereom (in grey; isotropic voxel size of 732 nm) and a simulated nodal approximation of the a single Diamond surface with a solid volume fraction $\phi=0.3$ (in yellow) and lattice parameter $a=39\, \mu{m}$ . Scale bar = $50\, \mu{m}$. The rectangular domain is oriented such that its orthogonal axes are the $[11\overline{2}]$, $[1\overline{1}0]$ and $[111]$ crystallographic directions of the Diamond geometry (in Miller index notation). (C) Slices through this subvolume showing the urchin and simulated data in the $[111]$, $[11\overline{2}]$, and $[1\overline{1}0]$ planes. Scale bar = $50\, \mu{m}$.} 
    \label{fig:Figure3}
\end{figure}

We also identified a stereom type that closely resembles the $Fd\overline{3}m$ Diamond geometry (Figure\ \ref{fig:Figure3}). This D-like stereom is located in a layer of stereom that lies in the centre of the plate below the P-like stereom. Remarkably, even large subvolumes containing more than 64 unit cells exhibit an almost perfect resemblance of the Diamond geometry (Figure\ \ref{fig:Figure3}B, C). Figure\ \ref{fig:Figure3}B displays a sample of the D-like stereom (in grey) such that the three orthogonal directions of the cube surrounding it align with the $[111]$, $[11\overline{2}]$, and $[1\overline{1}0]$ directions of the Diamond geometry. Adjacent to the D-like stereom in Figure\ \ref{fig:Figure3}B is a simulated nodal approximation of the Diamond geometry where the lattice parameter is equal to 39 microns and the solid volume fraction is 0.3. Figure\ \ref{fig:Figure3}C shows the close correspondence between cross-sections through each direction of the D-like stereom (in blue) and the simulated nodal approximation of the Diamond geometry (in grey). The lattice parameters, as estimated from the $\langle D_{CRT}\rangle$ value (Table\ \ref{tab:exp-PDG-area-curvature-results}), of the D-like stereom samples varied between 33 $\mu{m}$ and 39 $\mu{m}$. These estimates agree, up to approximately $\pm 1\, \mu{m}$, with estimates for the lattice parameter by manual registration of the nodal surface model. The structural properties of both the P-like and D-like stereom samples are discussed below and presented in Table\ \ref{tab:exp-PDG-area-curvature-results}.      

\begin{figure}[tbh]
    \centering
    \includegraphics[width=\textwidth]{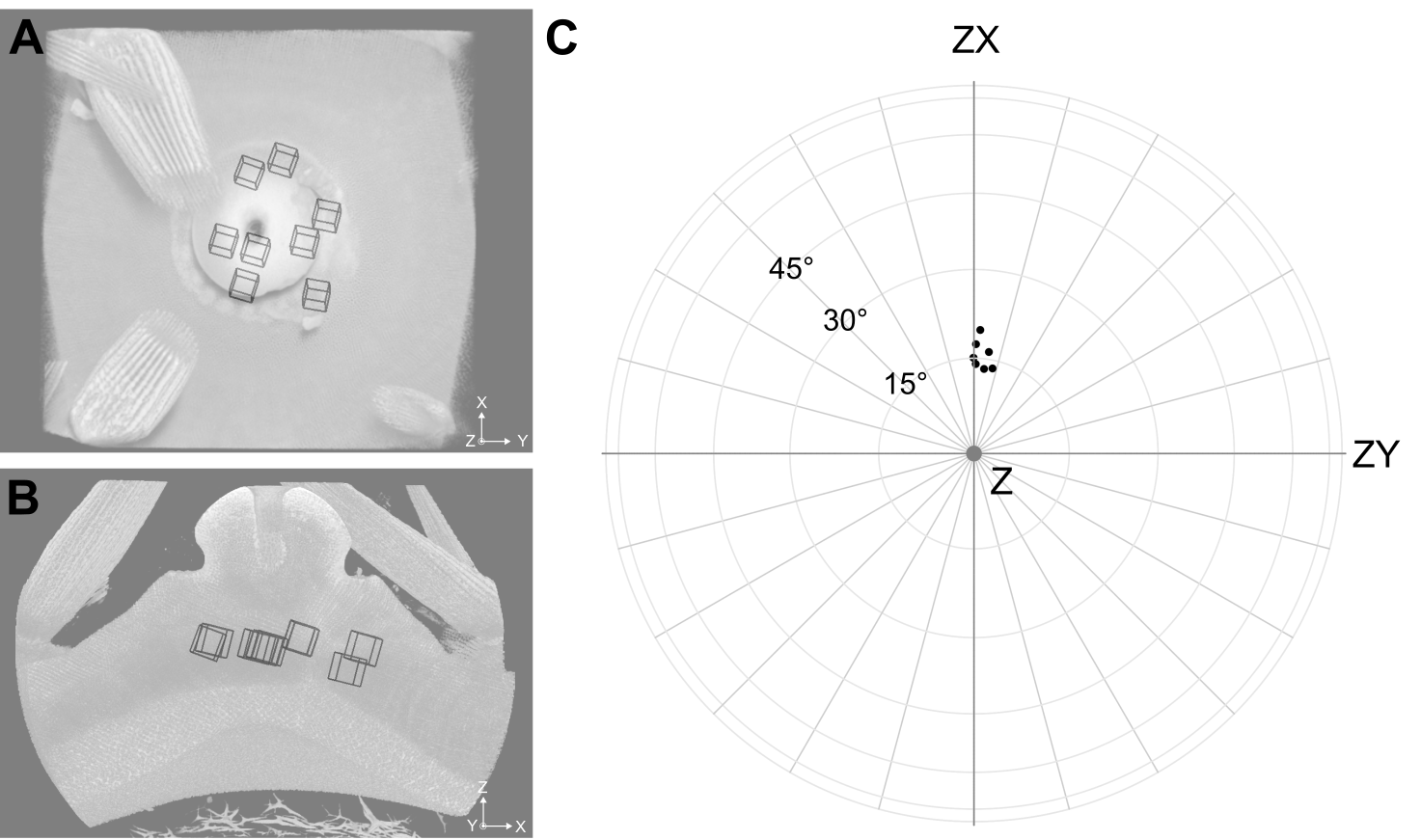}
    \caption{Orientation of the $[111]$ direction of the sea urchin D-like stereom relative to the direction of the spine (the positive Z-direction of the tomography dataset). (A, B) Volume render of the intermabulacral plate viewed along the Z-axis (A) and Y-axis (B) showing locations of eight sampled D-like stereom. (C) Orientation of the $[111]$ direction of each sample (black points) relative to the spine (grey point) represented by vectors projected onto a sphere. The sphere is presented in (C) as a perspective along the Z-axis. The mean orientation of the $[111]$ direction relative to the spine (the positive Z-direction) is 16$\pm$2 degrees.} 
    \label{fig:Figure4}
\end{figure}

Eight subvolumes of the D-like stereom were sampled across the interambulacral plate such that representative samples were taken across an approximately radial distribution centred on the tubercle (Figure\ \ref{fig:Figure4}A, B). The $[111]$ direction was easily identified in cross section and was shown to be almost uniformly oriented across all eight samples. All samples were oriented within 6 degrees from one another (Figure\ \ref{fig:Figure4}C). In relation to the spine (the positive Z-direction), the mean $[111]$ direction was oriented 16 $\pm$ 2 degrees approximately towards the positive X-direction (Figure\ \ref{fig:Figure4}C).

\subsection{Structural properties of ordered and disordered stereom types}

\begin{figure}[tbh]
    \centering
    \includegraphics[width=\textwidth]{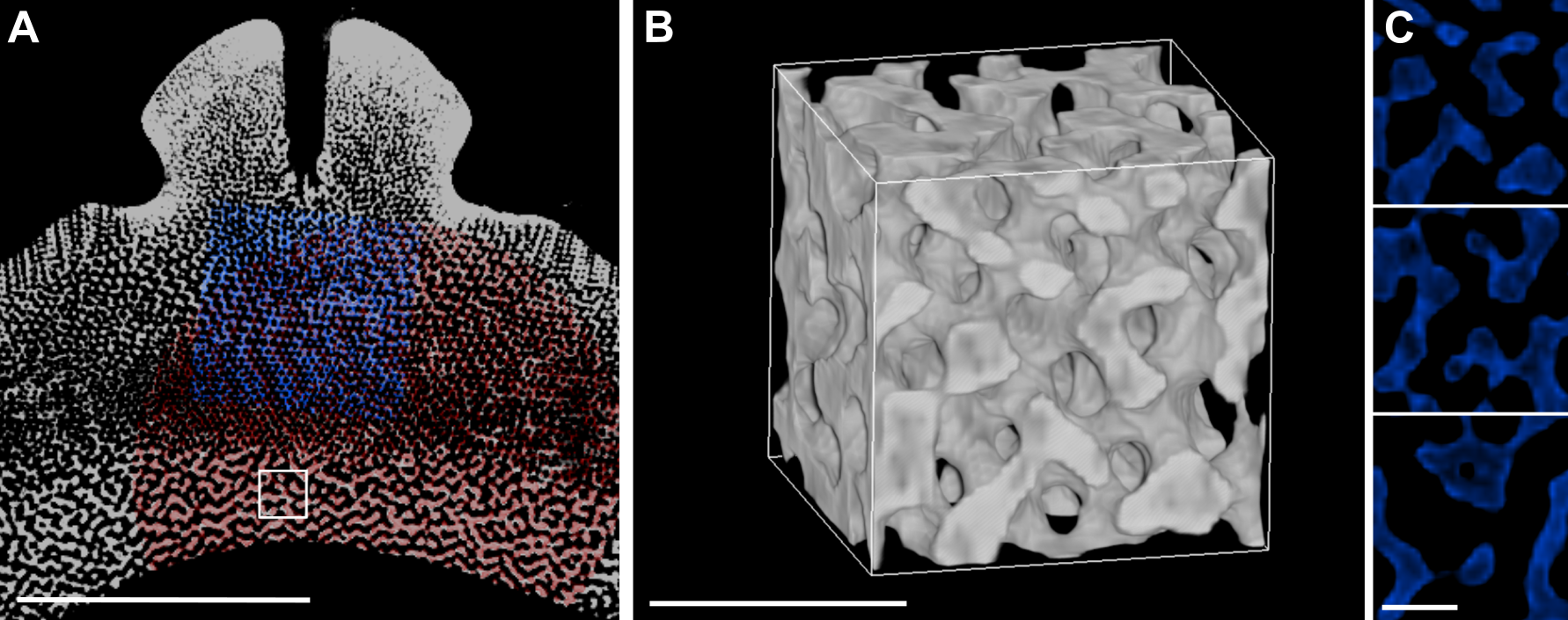}
    \caption{A representative subvolume of the Disordered stereom. (A) Cross-section through the interambulacral plate showing the location of the representative subvolume. Scale bar = 1 mm. (B) The subvolume of the sea urchin stereom (isotropic voxel size of $1.69\, \mu{m}$). Scale bar = $100\, \mu{m}$. (C) Perpendicular slices through the subvolume depicted in (B). Scale bar = $50\, \mu{m}$.} 
    \label{fig:Figure5}
\end{figure}

\begin{figure}[tbh]
    \centering
    \includegraphics[width=\textwidth]{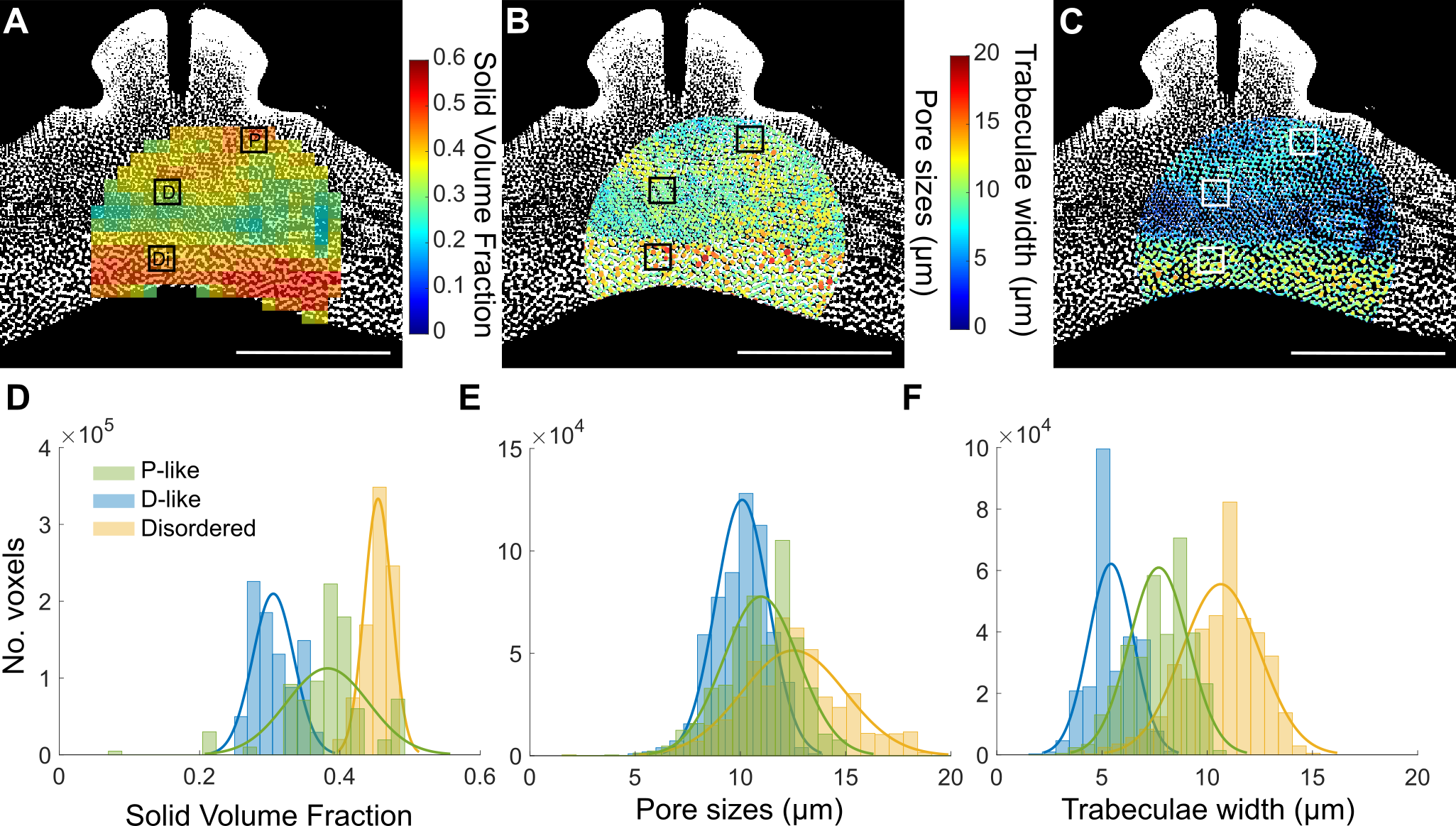}
    \caption{The distribution of solid volume fraction, pore sizes, and trabeculae widths across the interambulacral plate. (A) The distribution of solid volume fraction showing distinct differences across the plate. Scale bar = 1 mm. Squares P, D, and Di show locations of subsampled stereom where P is the P-like, D is the D-like, and Di is the Disordered stereom. (B) and (C) show the maximum pore sizes and trabeculae widths, respectively. (D), (E), and (F) show the distribution of solid volume fraction, pore sizes, and trabeculae widths across a subsample of each stereom type.}
    \label{fig:Figure6}
\end{figure}

In addition to the P-like and D-like ordered stereom types, a Disordered stereom type was identified and a representative sample is shown in Figure\ \ref{fig:Figure5}. The Disordered stereom type has an irregular network of trabeculae without pore alignment even in small subvolumes. It is found on the internal surface of the interambulacral plate and can easily be distinguished from other stereom types by its solid volume fraction ($0.48\pm0.03$; Figure\ \ref{fig:Figure6}A) and trabeculae widths ($11.0\pm0.4\, \mu$m; Figure\ \ref{fig:Figure6}C), that are approximately 30\% and 50\% larger than the D-like stereom located above it, respectively. This difference creates a sharp interface between Disordered and D-like stereom types as shown in Figure\ \ref{fig:Figure6}A and Figure\ \ref{fig:Figure6}C where no apparent gradual shift between the two occurs. 

Table\ \ref{tab:exp-PDG-area-curvature-results} summarizes the differences in structural properties between each of the identified stereom types. The solid volume fraction of the D-like stereom was relatively lower ($0.31\pm0.04$), compared to the P-like ($0.38\pm0.03$) and Disordered ($0.48\pm0.03$) stereom types. Figure\ \ref{fig:Figure6}D displays the distribution of solid volume fraction across each of the representative samples and shows the clear distinction between stereom types. 

The three stereom types have clear differences in the typical trabeculae width, that is, in the material thickness of the solid phase (Figure\ \ref{fig:Figure6}F). The D-like stereom has thinner solid elements ($\langle D_{CRT}^t\rangle =5.2\pm0.7$), compared to the P-like ($7.1\pm0.7$) and compared to the Disordered stereom ($11\pm0.4$). 

By contrast, the typical pore sizes within the three stereom types are much more uniform. The pore sizes are $\langle D_{CRT}^p\rangle=9.4\pm0.6\mu{m}$ for the D-like stereom, $10.3\pm0.9\mu{m}$ for the P-like stereom, and $11.8\pm0.6\mu{m}$ for the Disordered stereom. This is reflected also in the distributions of pore sizes for the representative samples shown in Figure\ \ref{fig:Figure6}E. We also show in the Supplementary Materials Figure S1, that this pore space percolates throughout the entire plate, connecting the external environment to the inside of the sea urchin. 

Compared to simulated nodal approximations (Supplementary Table S1), the ratio $\langle D_{CRT}^p\rangle/\langle D_{CRT}^t\rangle$ of pore sizes to trabeculae widths were larger in the sea urchin D-like and P-like stereom types. The ratio for the D-like stereom was between 6.8\% and 19.5\% larger than the simulated Diamond structure. The ratio for the P-like stereom was between 9.6\% and 12.7\% larger than the simulated Primitive structure. 

The interface areas of the trabeculae pore interfaces and their average mean curvatures were calculated using the Steiner method and are included in Table\ \ref{tab:exp-PDG-area-curvature-results}. There are small but significant differences between the stereom types that are consistent with the structural assignment of Diamond and Primitive geometries: The interface areas $A/V$ per volume are larger for the Diamond stereom (on average across the eight sampled domains, $A/V=0.10$) than for the P-like stereom (on average, $A/V=0.09$). The average mean curvatures are larger in absolute value for the D-like stereoms (on average $0.030$) than for the P-like stereom (on average $0.015$). These average values are consistent with the values for the exact Diamond and Primitive geometries in Table S1,  at the corresponding volume fractions and lattice parameters.

Compared to the ordered P-like and D-like stereom geometries, the disordered stereom has lower average mean curvature (on average across the eight sample domains $0.003$). That is, it is closer to a minimal surface, in line with the more balanced distribution of trabeculae (solid) material and pore space ($\phi\approx 0.48$).

\begin{table}[t]
  \caption{\label{tab:exp-PDG-area-curvature-results} \textbf{Key morphological properties of representative samples from the interambulacral plate of \textit{C. rugosa}.} Three samples from three types of stereom were examined including a D-like stereom, a P-like stereom, and a Disordered stereom. We provide estimates of the solid volume fraction $\phi$, the lattice parameter $a$, the maximal value of the Euclidean distance map $m_{EDM}$, the average $\langle D_{CRT}\rangle$ and the standard deviation of the maximal covering radius transform $\Delta D_{CRT}$, the surface area $A$ per volume, and the mean curvature $\langle H\rangle$ of the solid phase (representing the trabeculae) and pore (or void) phase of each sample. The superscripts 't' and 'p' indicate 'trabeculae' and 'pore', respectively. The volume of each sample is $0.004 mm^3$.}
          {\small

            \centering
  \begin{tabular}{l | l | l | l l l | l l l | l l }
\hline
& & &  \multicolumn{3}{c|}{solid phase (trabeculae)} & \multicolumn{3}{c|}{void phase (pore)} & &\\
\hline
Surface & $\phi$ & $a$ & $m_{EDM}^t$ & $\langle D_{CRT}^t\rangle$ & $\Delta D_{CRT}^t$ & $m_{EDM}^p$ & $\langle D_{CRT}^p\rangle$ & $\Delta D_{CRT}^p$ & $A/V$ & $\langle H\rangle$ \\
\hline
 & [1] & [$\mu{m}$] & [$\mu{m}$] & [$\mu{m}$] & [$\mu{m}$] & [$\mu{m}$] & [$\mu{m}$] & [$\mu{m}$] & [$\frac{1}{\mu{m}}$] & [$\frac{1}{\mu{m}}$] \\
\hline
D-like stereom 1 & 0.30 & 39 & 8.45 & 5.43 & 1.07 & 12.65 & 10.07 & 1.27 & 0.09 & 0.0209\\
D-like stereom 2 & 0.28 & 38 & 7.17 & 4.79 & 0.84 & 14.04 & 10.07 & 1.38 & 0.10 & 0.0376\\
D-like stereom 3 & 0.29 & 37 & 7.37 & 4.80 & 0.84 & 12.98 & 9.66 & 1.27 & 0.10 & 0.0484\\
D-like stereom 4 & 0.40 & 38 & 9.71 & 6.48 & 1.11 & 12.42 & 8.82 & 1.21 & 0.10 & 0.0158\\
D-like stereom 5 & 0.31 & 39 & 8.45 & 5.46 & 1.00 & 13.62 & 10.00 & 1.46 & 0.09 & 0.0252\\
D-like stereom 6 & 0.28 & 35 & 6.97 & 4.52 & 0.81 & 12.65 & 9.04 & 1.35 & 0.11 & 0.0509\\
D-like stereom 7 & 0.28 & 33 & 6.76 & 4.36 & 0.79 & 11.95 & 8.69 & 1.25 & 0.11 & 0.0277\\
D-like stereom 8 & 0.36 & 37 & 8.45 & 6.03 & 1.01 & 13.94 & 8.99 & 1.43 & 0.11 & 0.0090\\
P-like stereom 1 & 0.37 & 23 & 9.71 & 7.25 & 1.22 & 13.52 & 10.29 & 1.44 & 0.09 & 0.0199\\
P-like stereom 2 & 0.38 & 25 & 10.82 & 7.70 & 1.40 & 15.67 & 10.98 & 1.78 & 0.08  & 0.0200\\
P-like stereom 3 & 0.45 & 24 & 11.08 & 8.44 & 1.37 & 12.76 & 10.17 & 1.50 & 0.09  & 0.0056\\
P-like stereom 4 & 0.35 & 29 & 10.14 & 6.98 & 1.28 & 16.30 & 11.63 & 2.08 & 0.08  & 0.0170\\
P-like stereom 5 & 0.35 & 27 & 9.85 & 6.82 & 1.22 & 14.44 & 10.76 & 1.55 & 0.09  & 0.0219\\
P-like stereom 6 & 0.36 & 25 & 9.26 & 6.68 & 1.21 & 13.62 & 10.35 & 1.51 & 0.09  & 0.0243\\
P-like stereom 7 & 0.40 & 28 & 10.01 & 6.31 & 1.17 & 13.62 & 8.80 & 1.57 & 0.10  & 0.0062\\
P-like stereom 8 & 0.41 & 27 & 10.42 & 6.91 & 1.32 & 14.54 & 9.31 & 1.72 & 0.11  & 0.0058\\
Disord.\ stereom 1 & 0.45 & - & 14.64 & 10.64 & 1.85 &  17.96 & 12.53 & 2.45 & 0.07 & 0.0062\\
Disord.\ stereom 2 & 0.44 & - & 14.64 & 10.41 & 1.77 & 17.48 & 12.42 & 2.24 & 0.07 & 0.0049\\
Disord.\ stereom 3 & 0.47 & - & 14.64 & 10.80 & 1.96 & 16.39 & 12.27 & 2.28 & 0.07 & 0.0080\\
Disord. stereom 4 & 0.51 & - & 15.21 & 11.43 & 1.84 & 15.48 & 11.10 & 2.19 & 0.07 & 0.0003\\
Disord. stereom 5 & 0.47 & - & 14.93 & 10.91 & 2.00 & 18.89 & 11.17 & 2.36 & 0.07 & 0.0024\\
Disord. stereom 6 & 0.48 & - & 15.30 & 11.19 & 1.96 & 14.83 & 11.04 & 1.91 & 0.07 & 0.0013\\
Disord. stereom 7 & 0.49 & - & 15.11 & 10.94 & 1.87 & 16.81 & 11.75 & 2.37 & 0.07 & 0.0005\\
Disord. stereom 8 & 0.51 & - & 16.29 & 11.47 & 1.94 & 16.98 & 11.80 & 2.31 & 0.07 & 0.0039\\ \hline
  \end{tabular}
  }
\end{table}

\subsection{X-ray Scattering of the Disordered and Primitive stereom}

\begin{figure}[t]
    \centering
    \includegraphics[width=\textwidth]{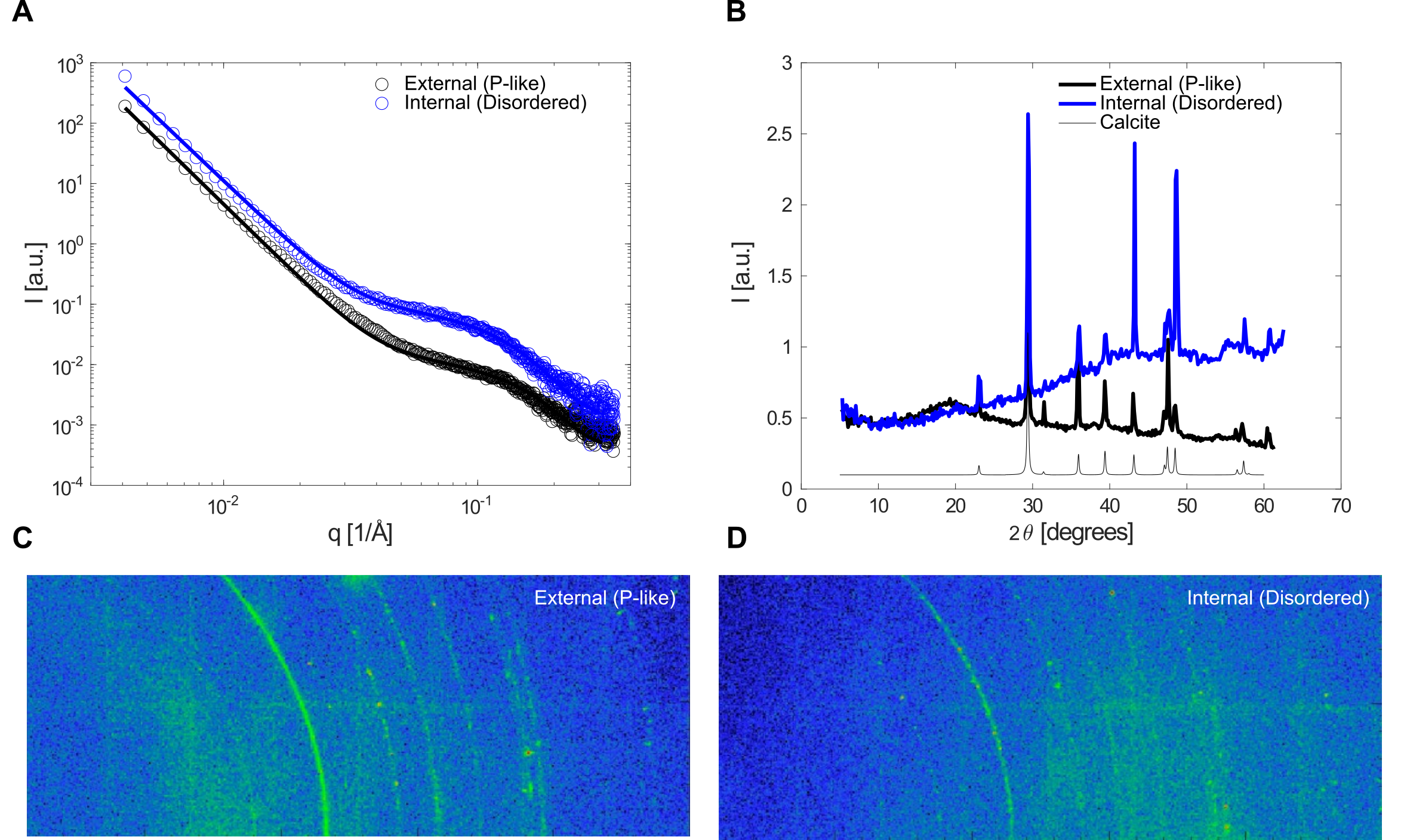}
    \caption{Azimuthally averaged intensities of SAXS (A) and WAXS (B) data from internal and external regions of the stereom that are characterised by micron-scale "Internal (Disordered)" and "External (P-like)" surface structures, respectively. The data labelled "Calcite" is reference data for calcite \cite{sitepu2005comparative}. (C,D) 2D detector images of and WAXS for the "External (P-like)" and "Internal (Disordered)" samples.}
    \label{fig:Figure7}
\end{figure}

The structural X-ray tomography analyses at the micron-scale are complemented with some preliminary SAXS and WAXS analyses of the \AA ngstrom and nanometer scale structure within the constituent material. We analysed two samples representing different regions of the sea urchin stereom. One of these regions is on the internal surface of the plate where we expect the porous structure to be disordered; this sample is labelled "Internal (Disordered)". The other region is closer to the external surface in a position where we expect the microstructure to be the Primitive surface; this sample is labelled "External (P-like)". Figure \ref{fig:Figure7}A and B show the SAXS and WAXS results, respectively. 

For both samples, the 2D scattering intensity exhibits a pattern that is not consistent with a single crystal structure (discrete spots), but rather with a polycrystalline structure of randomly oriented crystallites (continuous concentric circles $I(\mathbf{q})=I(|\mathbf{q}|)$); we attribute the remnant granularity to the polycrystalline nature with relatively large crystallites.\footnote{Due to the experimental setup the 2D detector only register approximately a fifth of the full pattern, thus any larger crystallites in the ‘right’ orientation will contribute disproportionately if registered - we believe this to be the reason for the different intensities between the two data sets, for example the [2,0,2] reflection at 43.1 degrees.} 

From the WAXS data in Figure\ \ref{fig:Figure7}B we see that both samples have the same Bragg-peaks, consistent with the crystal structure of calcite \cite{sitepu2005comparative}.  

However, the sample from the internal region where the stereom has a disordered micron-scale microstructure has a clear underlying amorphous contribution for 2$\theta$ angles over approximately 20 degrees that must stem from a material with an amorphous structure at the molecular (nanometer) scale. It is tempting to speculate that the amorphous scattering could be a signal that stems from an amorphous calcium carbonate phase present in the native sea urchin. Given that sample preparation in regard to the ordered and the disordered regions was the same, it is unlikely but not impossible that the amorphous signal stems from sample preparation artefacts (e.g.\ mechanochemical effects such as grinding).

We attribute the difference in nano-domain size deduced from the SAXS data to this molecular scale difference. A fit with a model combining a low-q power law with a Unified scattering model term \cite{beaucage20122} allows the determination of a structural measure in the form of a radius of gyration which can be interpreted as an average domain size in the calcite. We find that the average domain size is larger in the Disordered stereom compared to the ordered P-like stereom. The radii of gyration are 1.57 nm and 1.32 nm for the Internal sample (Disordered stereom) and External sample (P-like stereom), respectively.

\section{Discussion and conclusions}

By combining high-resolution three-dimensional data with structural modelling and mapping, we have here, for the first time, identified the co-occurence of micron-scale Diamond and Primitive TPMS-related structures along with a disordered structure in the endoskeleton of a sea urchin (Figure\ \ref{fig:Figure2}, \ref{fig:Figure3}, \ref{fig:Figure5}). Structural properties including solid volume fraction and trabeculae widths differ to the extent that contrasting domain boundaries are clearly visible (Figure\ \ref{fig:Figure6}). Additionally, small differences in interface area and mean curvature were also found between the three bicontinuous structures (Table\ \ref{tab:exp-PDG-area-curvature-results}). The pore sizes across the plate remain relatively consistent and we show that the void space of the plate is a single connected component, connecting the external environment with the inside of the sea urchin (Figure S1).

Our structural analyses have established both differences and similarities between the three identified bicontinuous geometries (D-like, P-like and Disordered). A standout feature is that the pore sizes in all three structures are fairly uniform, whereas both the solid volume fraction and the trabeculae widths vary considerably. 

In terms of the network topology of the disordered structure, it remains to be seen whether and how it relates to the ordered structures, but observations in smaller scale biological photonic crystals have shown that the amorphous networks in weevil scales are related to their ordered counterparts \cite{djeghdi20223d, bauernfeind2023photonic}. 

While our X-ray scattering analyses are preliminary, they allow us to consider structural models for the relationship between the Disordered and P-like stereom. There is no difference between the calcite crystallite sizes, as extracted from WAXS, but there is a difference in the radius of gyration, as extracted from SAXS. A structural hypothesis, could thus be that the Disordered stereom is a mixture of calcite crystallites dispersed in an amorphous matrix leading to a larger effective domain size of combined crystalline/amorphous material. If the increased nano-domains are interpreted as a ‘swelling’ of the crystal by amorphous calcium carbonate, this may explain the observed larger trabeculae widths and solid volume fractions of the Disordered stereom; the increase in dimensions of the crystallites (as determined by SAXS) is approximately 19\% (1.57 nm/1.32 nm) whereas the increase in trabeculae width (as determined by tomography) is approximately 35\% ($10.5\, \mu{m}/7.7\, \mu{m})$. Further investigations, through synchrotron analysis where greater flux enables a spatially more fine-grained analysis including on thicker samples, could test this hypothesis.

The unique distribution of ordered (Diamond and Primitive) and disordered bicontinuous structures in \textit{C. rugosa} offers an opportunity to examine the possible functions of different stereom types, aside from their obvious role in mechanical support \cite{yang2022damage}. In living urchins, the spines are attached to the tubercles via a ball-and-socket joint that is connected to the plate by collagenous fibres that penetrate the long galleries of the surface stereom (that Smith refers to as 'rectilinear' stereom) \cite{smith1980stereom}. As Smith (1980) highlighted in his comprehensive study, the type of investing soft tissue is clearly related to the morphology of the stereom and perhaps even influences its growth. Smith (1980) reported that the 'rectilinear' stereom found in cidaroids was associated with a diverse set of soft tissue including ectoderm, muscle fibres, and collagenous fibres, whereas the 'labyrinthic' stereom (Disordered stereom in our study) was associated with an endothelial layer. Our results now show that this 'rectilinear' stereom, while made up of ordered bicontinuous structures, does include different types of ordered structures with different structural properties. It would therefore be very informative to the functional interpretation of these structures to examine the specific types of investing soft tissue that are associated with the D-like and P-like stereom types, respectively.

While the Primitive surface related structure reported here was observed many decades ago \cite{Nissen1969-ph}, the occurrence of the Diamond surface related structure is novel and represents only the third occurrence of this structure at a micron-sized length scale \cite{yang2022damage, gorzelak2023devonian}. Despite there being very few examples of ordered bicontinuous structures at such large length scales, the recent discoveries of Diamond surface related structures in the knobby starfish and a fossilized crinoid highlight that this structure is found across distantly related clades \cite{yang2022damage, gorzelak2023devonian}. Gorzelak et al. (2023) suggest that this may indicate the independent evolution of this structure between echinoderm clades or alternatively may indicate a more widespread genomically encoded trait that emerges during increased predation/mechanical stress periods. Indeed, one of the unique properties of this Diamond structure is enhanced damage tolerance \cite{yang2022damage}, however whether this property evolved for this purpose or is merely a by-product of the formation process is yet to be seen.

The biogenesis of the ordered bicontinuous structures remains a key question, despite substantive progress in understanding the biomineralisation mechanisms. We share the view expressed by Hyde \& Meldrum \cite{Hyde2022-xo} that the "ultimate question of what determines the ultrastructure remains open" and that we can "not explain how those structures can form with huge lattice parameters ($>10\, \mu{m}$)". 

One aspect concerns the formation of smooth interfaces in a material usually associated with sharp crystalline features (i.e., geogenic calicte). It is now accepted that rather than being true single crystals, the biogenic calcite of echinoderm stereom is better described as 'mesocrystalline' comprising small ($0.1-0.3\, \mu{m}$) domains of co-oriented crystalline calcite and it is thought that this property allows the crystal lattice to form curved surfaces \cite{Hyde2022-xo,smith1990biomineralization}. It is believed that the mesocrystalline structure is achieved as a consequence of the crystallization pathway whereby calcite forms via transient amorphous precursor phases involving nanoparticles of amorphous calcium carbonate (ACC) \cite{Hyde2022-xo}. 

Another fascinating aspect of the formation of calcite structures in sea urchins and other marine organisms is the high degree of alignment of the c-axis of calcite with both the macroscopic features of the sea urchin (such as the direction of the spines \cite{Killian2009}) and with the orientation of the micron-scale ordered porous structures (such as the reported alignment of calcite c-axis with the [111] direction of the micron-scale Diamond structure in a starfish \cite{yang2022damage}). We have here reported that the orientation of the [111] direction of the D-like stereom is uniform across the interambulacral plate, at a constant angle of about 15$^o$ to the spine direction. We have not been able to identify any other obvious morphological feature (such as the skeleton surface normal) that this [111] direction aligns with. Whether the calcite c-axis aligns with the [111] direction of our D-like stereom, such as is the case in the starfish, is yet to be seen.

The even more intriguing question is, in our view, how the formation of highly-ordered Diamond-like or Primitive-like structures takes place at the very large length scales of microns. These length scales are so large that the entropic or enthalpic mechanisms responsible for the formation of related structures in lipid (bilayer) phases -- at much smaller scales -- are irrelevant and cannot drive this structure formation. Yet, closely related structures form. Skeletogenesis in sea urchins has been a well studied topic, however almost all studies have focused on skeletogenesis in larval sea urchins not in the adult forms where we find these highly-ordered large-scale structures. Nonetheless, skeletogenesis in sea urchin larvae can certainly inform some aspects of how these structures form. For example, it is well understood that the larval skeleton forms within a syncytium, a thin cytoplasmic cable (3 $\mu$m in diameter) formed by many connected skeletogenic cells \cite{beniash1999cellular, wilt1999matrix, wilt2007morphogenesis,politi2008transformation}. The ultrastructural morphology of the larval skeleton is therefore dictated by the organic matrix that forms the syncytium. Over the last two decades it has become apparent that certain signalling molecules, namely vascular endothelial growth factor, play an integral role in skeletogenic cell patterning and gene expression, syncytium formation, and therefore in controlling the gross geometry of the larval echinoderm skeleton \cite{duloquin2007localized, knapp2012recombinant, adomako2013growth, sun2014signal, morgulis2021vegf, tarsis2022distinct}. Many of the same molecular and cellular processes directing skeletogenesis in larval echinoderms also play a role in juvenile and adult skeletogenesis \cite{thompson2021post}, therefore it is likely that the very same signalling molecules are influencing the development of the adult ultrastructural forms.

It is clear that the formation process of the bicontinuous geometries in the sea urchin (and other echinoderms) is substantially different from the formation of bicontinuous lipid membrane forms -- the length scale is hugely different and it does not involve a nanoscale intracellular membrane sculpted to a bicontinuous form (the role of the shape of the plasma membrane of the syncytium remains an open question). It seems likely that the formation mechanism will involve some elements that can be explained by generic mathematical pattern formation processes, some elements that will relate to calcium carbonate biomineralisation, some elements that will be specific to the biochemistry, and some elements that relate to the evolutionary benefits of the microstructural forms. It will be exciting to see how tomography studies, such as are conducted here, of mature and developing sea urchins will help to elucidate both the mathematical, physical, chemical, and biological elements of this process.


\section*{Acknowledgements}
All data in the article are available on Dryad (https://doi.org/10.5061/dryad.wdbrv15vf).

We gratefully acknowledge funding by the Australian Research Council (ARC) through the Discovery Project DP200102593. A.J.M is supported by an Australian Government Research Training Program (RTP) Scholarship.

We thank Dr Charles Messing (Nova Southeastern University of Florida) for a donation of the {\it C.\ rugosa} specimen. We thank Rasmus Klarskov Nordal E Petersen for help with X-ray sample preparation. We are grateful to Prof Bodo Wilts and the three anonymous reviewers for insightful comments and helpful suggestions on the mansucript. The authors acknowledge the facilities of Microscopy Australia at the Centre for Microscopy, Characterisation and Analysis at The University of Western Australia.



\bibliography{refs} 

\begin{thebibliography}{10}

\bibitem{adomako2013growth}
Ashrifia Adomako-Ankomah and Charles~A Ettensohn.
\newblock Growth factor-mediated mesodermal cell guidance and skeletogenesis
  during sea urchin gastrulation.
\newblock {\em Development}, 140(20):4214--4225, 2013.

\bibitem{ALMSHERQI2009275}
Zakaria Almsherqi, Tomas Landh, Sepp Kohlwein, and Yuru Deng.
\newblock Chapter 6 cubic membranes: The missing dimension of cell membrane
  organization.
\newblock In {\em International Review of Cell and Molecular Biology}, volume
  274 of {\em International Review of Cell and Molecular Biology}, pages
  275--342. Academic Press, 2009.

\bibitem{bauernfeind2023photonic}
Viola Bauernfeind, Kenza Djeghdi, Ilja Gunkel, Ullrich Steiner, and Bodo Wilts.
\newblock Photonic amorphous i-wp-like networks create angle-independent colors
  in sternotomis virescens longhorn beetles.
\newblock {\em Advanced Functional Materials}, page 2302720, 2023.

\bibitem{beaucage20122}
G~Beaucage.
\newblock 2.14-combined small-angle scattering for characterization of
  hierarchically structured polymer systems over nano-to-micron meter: Part ii
  theory,”.
\newblock {\em Polymer science: a comprehensive reference. Elsevier,
  Amsterdam}, 399, 2012.

\bibitem{beniash1999cellular}
Elia Beniash, Lia Addadi, and Stephen Weiner.
\newblock Cellular control over spicule formation in sea urchin embryos: A
  structural approach.
\newblock {\em Journal of structural biology}, 125(1):50--62, 1999.

\bibitem{djeghdi20223d}
Kenza Djeghdi, Ullrich Steiner, and Bodo Wilts.
\newblock 3d tomographic analysis of the order-disorder interplay in the
  pachyrhynchus congestus mirabilis weevil.
\newblock {\em Advanced Science}, 9(26):2202145, 2022.

\bibitem{donnay1969x}
Gabrielle Donnay and David~L Pawson.
\newblock X-ray diffraction studies of echinoderm plates.
\newblock {\em Science}, 166(3909):1147--1150, 1969.

\bibitem{dove2018biomineralization}
Patricia~M Dove, James~J De~Yoreo, and Steve Weiner.
\newblock {\em Biomineralization}, volume~54.
\newblock Walter de Gruyter GmbH \& Co KG, 2018.

\bibitem{duloquin2007localized}
Louise Duloquin, Guy Lhomond, and Christian Gache.
\newblock Localized vegf signaling from ectoderm to mesenchyme cells controls
  morphogenesis of the sea urchin embryo skeleton.
\newblock 2007.

\bibitem{gorzelak2021functional}
Przemyslaw Gorzelak.
\newblock {\em Functional Micromorphology of the Echinoderm Skeleton}.
\newblock Cambridge University Press, 2021.

\bibitem{gorzelak2023devonian}
Przemys{\l}aw Gorzelak, Dorota Ko{\l}buk, Jaros{\l}aw Stolarski, Pawe{\l}
  B{k{a}}cal, Bart{\l}omiej Januszewicz, Piotr Duda, Dorota {\'S}rodek, Tomasz
  Brachaniec, and Mariusz Salamon.
\newblock A devonian crinoid with a diamond microlattice.
\newblock {\em Proceedings of the Royal Society B}, 290(1995):20230092, 2023.

\bibitem{ha2004three}
Y-H Ha, Richard~A Vaia, William~F Lynn, Joseph~P Costantino, Jennifer Shin,
  Andrew~B Smith, Paul~T Matsudaira, and Edwin~L Thomas.
\newblock Three-dimensional network photonic crystals via cyclic size
  reduction/infiltration of sea urchin exoskeleton.
\newblock {\em Advanced Materials}, 16(13):1091--1094, 2004.

\bibitem{hain2023spire}
Tobias~M Hain, Michal Bykowski, Matthias Saba, Myfanwy~E Evans, Gerd~E
  Schr\"oder-Turk, and Lucja Kowalewska.
\newblock {SPIRE—a software tool for bicontinuous phase recognition:
  application for plastid cubic membranes}.
\newblock {\em Plant Physiology}, 188(1):81--96, 10 2021.

\bibitem{han2018overview}
Lu~Han and Shunai Che.
\newblock An overview of materials with triply periodic minimal surfaces and
  related geometry: from biological structures to self-assembled systems.
\newblock {\em Advanced Materials}, 30(17):1705708, 2018.

\bibitem{Hyde2009Elusive}
Stephen Hyde, Michael O'Keeffe, and Davide Proserpio.
\newblock A short history of an elusive yet ubiquitous structure in chemistry,
  materials, and mathematics.
\newblock {\em Angewandte Chemie International Edition}, 47(42):7996--8000,
  2008.

\bibitem{Hyde2022-xo}
Stephen~T Hyde and Fiona~C Meldrum.
\newblock Starfish grow extraordinary crystals.
\newblock {\em Science}, 375(6581):615--616, February 2022.

\bibitem{Killian2009}
Christopher~E. Killian, Rebecca~A. Metzler, Y.~U.~T. Gong, Ian~C. Olson, Joanna
  Aizenberg, Yael Politi, Fred~H. Wilt, Andreas Scholl, Anthony Young, Andrew
  Doran, Martin Kunz, Nobumichi Tamura, Susan~N. Coppersmith, and P.~U. P.~A.
  Gilbert.
\newblock Mechanism of calcite co-orientation in the sea urchin tooth.
\newblock {\em Journal of the American Chemical Society}, 131(51):18404--18409,
  2009.
\newblock PMID: 19954232.

\bibitem{knapp2012recombinant}
Regina~T Knapp, Ching-Hsuan Wu, Kellen~C Mobilia, and Derk Joester.
\newblock Recombinant sea urchin vascular endothelial growth factor directs
  single-crystal growth and branching in vitro.
\newblock {\em Journal of the American Chemical Society}, 134(43):17908--17911,
  2012.

\bibitem{Kowalewska2019review}
Lucja Kowalewska, Michal Bykowski, and Agnieszka Mostowska.
\newblock Spatial organization of thylakoid network in higher plants.
\newblock {\em Botany Letters}, 166(3):326--343, 2019.

\bibitem{LUZZATI1997661}
Vittorio Luzzati.
\newblock Biological significance of lipid polymorphism: the cubic phases.
\newblock {\em Current Opinion in Structural Biology}, 7(5):661--668, 1997.

\bibitem{LUZZATI200417}
Vittorio Luzzati, Hervé Delacroix, Annette Gulik, Tadeusz Gulik-Krzywicki,
  Paolo Mariani, and Rodolfo Vargas.
\newblock The cubic phases of lipids.
\newblock In Osamu Terasaki, editor, {\em Mesoporous Crystals and Related
  Nano-Structured Materials}, volume 148 of {\em Studies in Surface Science and
  Catalysis}, pages 17--40. Elsevier, 2004.

\bibitem{MezzengaAdvMater2019}
Raffaele Mezzenga, John Seddon, Calum Drummond, Ben Boyd, Gerd Schröder-Turk,
  and Laurent Sagalowicz.
\newblock Nature-inspired design and application of lipidic lyotropic liquid
  crystals.
\newblock {\em Advanced Materials}, 31(35):1900818, 2019.

\bibitem{michielsen2008gyroid}
Kristel Michielsen and Doekele Stavenga.
\newblock Gyroid cuticular structures in butterfly wing scales: biological
  photonic crystals.
\newblock {\em Journal of The Royal Society Interface}, 5(18):85--94, 2008.

\bibitem{Mickel2008}
Walter Mickel, Stefan M\"{u}nster, Louise~M. Jawerth, David~A. Vader, David~A.
  Weitz, Adrian~P. Sheppard, Klaus Mecke, Ben Fabry, and Gerd~E.
  Schr\"{o}der-Turk.
\newblock Robust pore size analysis of filamentous networks from
  three-dimensional confocal microscopy.
\newblock {\em Biophysical Journal}, 95(12):6072--6080, December 2008.

\bibitem{morgulis2021vegf}
Miri Morgulis, Mark~R Winter, Ligal Shternhell, Tsvia Gildor, and Smadar
  Ben-Tabou de~Leon.
\newblock Vegf signaling activates the matrix metalloproteinases, mmpl7 and
  mmpl5 at the sites of active skeletal growth and mmpl7 regulates skeletal
  elongation.
\newblock {\em Developmental Biology}, 473:80--89, 2021.

\bibitem{Nissen1969-ph}
H~U Nissen.
\newblock Crystal orientation and plate structure in echinoid skeletal units.
\newblock {\em Science}, 166(3909):1150--1152, November 1969.

\bibitem{perricone2023echinoid}
Valentina Perricone, Pasquale Cesarano, Andrea Mancosu, Davide Asnicar, Sergio
  Bravi, and Francesco Marmo.
\newblock Echinoid skeleton: an insight on the species-specific pattern of the
  paracentrotus lividus plate and its microstructural variability.
\newblock {\em Journal of the Royal Society Interface}, 20(199):20220673, 2023.

\bibitem{perricone2022hexagonal}
Valentina Perricone, Tobias~B Grun, Francesco Rendina, Francesco Marmo,
  Maria~Daniela Candia~Carnevali, Michal Kowalewski, Angelo Facchini, Mario
  De~Stefano, Luigia Santella, Carla Langella, et~al.
\newblock Hexagonal voronoi pattern detected in the microstructural design of
  the echinoid skeleton.
\newblock {\em Journal of the Royal Society Interface}, 19(193):20220226, 2022.

\bibitem{politi2008transformation}
Yael Politi, Rebecca~A Metzler, Mike Abrecht, Benjamin Gilbert, Fred~H Wilt,
  Irit Sagi, Lia Addadi, Steve Weiner, and PUPA Gilbert.
\newblock Transformation mechanism of amorphous calcium carbonate into calcite
  in the sea urchin larval spicule.
\newblock {\em Proceedings of the National Academy of Sciences},
  105(45):17362--17366, 2008.

\bibitem{SaranathanBirdsPNAS}
Vinodkumar Saranathan, Suresh Narayanan, Alec Sandy, Eric Dufresne, and Richard
  Prum.
\newblock Evolution of single gyroid photonic crystals in bird feathers.
\newblock {\em Proceedings of the National Academy of Sciences},
  118(23):e2101357118, 2021.

\bibitem{saranathan2010structure}
Vinodkumar Saranathan, Chinedum Osuji, Simon Mochrie, Heeso Noh, Suresh
  Narayanan, Alec Sandy, Eric Dufresne, and Richard Prum.
\newblock Structure, function, and self-assembly of single network gyroid (i
  4132) photonic crystals in butterfly wing scales.
\newblock {\em Proceedings of the National Academy of Sciences},
  107(26):11676--11681, 2010.

\bibitem{sitepu2005comparative}
Husin Sitepu, Brian O'Connor, and Deyu Li.
\newblock Comparative evaluation of the march and generalized spherical
  harmonic preferred orientation models using x-ray diffraction data for
  molybdite and calcite powders.
\newblock {\em Journal of Applied Crystallography}, 38(1):158--167, 2005.

\bibitem{smith1990biomineralization}
Andrew Smith and J~Carter.
\newblock Biomineralization in echinoderms.
\newblock {\em Skeletal biomineralization: patterns, processes and evolutionary
  trends}, 1:413--442, 1990.

\bibitem{smith1980stereom}
Andrew~B Smith.
\newblock Stereom microstructure of the echinoid test.
\newblock 1980.

\bibitem{sun2014signal}
Zhongling Sun and Charles~A Ettensohn.
\newblock Signal-dependent regulation of the sea urchin skeletogenic gene
  regulatory network.
\newblock {\em Gene Expression Patterns}, 16(2):93--103, 2014.

\bibitem{takahashi1967ball}
Keiichi Takahashi.
\newblock The ball-and-socket joint of the sea-urchin spine: geometry and its
  functional implications.
\newblock {\em Journal of the Faculty of Science, University of Tokyo.},
  11:131--135, 1967.

\bibitem{tarsis2022distinct}
Kristina Tarsis, Tsvia Gildor, Miri Morgulis, and Smadar Ben-Tabou~de Leon.
\newblock Distinct regulatory states control the elongation of individual
  skeletal rods in the sea urchin embryo.
\newblock {\em Developmental Dynamics}, 251(8):1322--1339, 2022.

\bibitem{thompson2021post}
Jeffrey Thompson, Periklis Paganos, Giovanna Benvenuto, Maria~Ina Arnone, and
  Paola Oliveri.
\newblock Post-metamorphic skeletal growth in the sea urchin paracentrotus
  lividus and implications for body plan evolution.
\newblock {\em EvoDevo}, 12:1--14, 2021.

\bibitem{Thovert}
J.-F. Thovert, F.~Yousefian, P.~Spanne, C.~G. Jacquin, and P.~M. Adler.
\newblock Grain reconstruction of porous media: Application to a low-porosity
  fontainebleau sandstone.
\newblock {\em Phys. Rev. E}, 63:061307, May 2001.

\bibitem{tian2023sea}
Yang Tian, Xianglei Liu, Qingyang Luo, Haichen Yao, Jianguo Wang, Chunzhuo
  Dang, Shushan Lv, Qiao Xu, Jiawei Li, Li~Zhang, et~al.
\newblock Sea urchin skeleton-inspired triply periodic foams for fast latent
  heat storage.
\newblock {\em International Journal of Heat and Mass Transfer}, 206:123944,
  2023.

\bibitem{weiner2008biomineralization}
Stephen Weiner.
\newblock Biomineralization: a structural perspective.
\newblock {\em Journal of structural biology}, 163(3):229--234, 2008.

\bibitem{wilt1999matrix}
Fred~H Wilt.
\newblock Matrix and mineral in the sea urchin larval skeleton.
\newblock {\em Journal of structural biology}, 126(3):216--226, 1999.

\bibitem{wilt2007morphogenesis}
Fred~H Wilt and Charles~A Ettensohn.
\newblock The morphogenesis and biomineralization of the sea urchin larval
  skeleton.
\newblock {\em Handbook of biomineralization: biological aspects and structure
  formation}, pages 182--210, 2007.

\bibitem{wilts2012brilliant}
Bodo Wilts, Kristel Michielsen, Jeroen Kuipers, Hans De~Raedt, and Doekele
  Stavenga.
\newblock Brilliant camouflage: photonic crystals in the diamond weevil,
  entimus imperialis.
\newblock {\em Proceedings of the Royal Society B: Biological Sciences},
  279(1738):2524--2530, 2012.

\bibitem{WiltsScienceAdv2017}
Bodo~D. Wilts, Benjamin~Apeleo Zubiri, Michael~A. Klatt, Benjamin Butz,
  Michael~G. Fischer, Stephen~T. Kelly, Erdmann Spiecker, Ullrich Steiner, and
  Gerd~E. Schröder-Turk.
\newblock Butterfly gyroid nanostructures as a time-frozen glimpse of
  intracellular membrane development.
\newblock {\em Science Advances}, 3(4):e1603119, 2017.

\bibitem{yang2022damage}
Ting Yang, Hongshun Chen, Zian Jia, Zhifei Deng, Liuni Chen, Emily~M Peterman,
  James~C Weaver, and Ling Li.
\newblock A damage-tolerant, dual-scale, single-crystalline microlattice in the
  knobby starfish, protoreaster nodosus.
\newblock {\em Science}, 375(6581):647--652, 2022.

\bibitem{yang2022high}
Ting Yang, Zian Jia, Ziling Wu, Hongshun Chen, Zhifei Deng, Liuni Chen, Yunhui
  Zhu, and Ling Li.
\newblock High strength and damage-tolerance in echinoderm stereom as a natural
  bicontinuous ceramic cellular solid.
\newblock {\em Nature Communications}, 13(1):6083, 2022.

\end{thebibliography}
\bibliographystyle{plain} 

\end{document}